\documentclass[twocolumn,showpacs,prd]{revtex4}
\usepackage{graphicx}% Include figure files
\usepackage{dcolumn}% Align table columns on decimal point
\usepackage{bm}% bold math

\begin{document}

 \preprint{hep-th/0204059}

\title{Chern-Simons Term for BF Theory and
Gravity as a Generalized \\Topological Field Theory in Four
Dimensions}

\author{Han-Ying Guo${}^{1,2}$}
\email{hyguo@itp.ac.cn}

\author{Yi Ling${}^{2,3}$}
\email{ling@phys.psu.edu}

\author{Roh-Suan Tung${}^{2,4}$}
\email{tung@dirac.ucdavis.edu}

\author{Yuan-Zhong Zhang${}^{1,2}$}
\email{yzhang@itp.ac.cn}

\affiliation{%
${}^1$ CCAST (World Laboratory), P.O. Box 8730, Beijing
   100080, China}

\affiliation{%
${}^2$ Institute of Theoretical Physics,
 Chinese Academy of Sciences
 P.O.Box 2735, Beijing 100080, China}

\affiliation{%
${}^3$ Center for Gravitational Physics and
Geometry, Department of Physics, Pennsylvania
State University, University Park, PA 16802, USA}

\affiliation{%
${}^4$ Department of Physics, University of
California, Davis, CA 95616, USA}

\date{\today}

\begin{abstract}
\quad A direct relation between two types of topological field
theories, Chern-Simons theory and $BF$ theory, is presented by
using ``Generalized Differential Calculus'', which extends an
ordinary $p-$form to an ordered pair of $p$ and $(p+1)-$form. We
first establish the generalized Chern-Weil homomormism for
generalized curvature invariant polynomials in general even
dimensional manifolds, and then show that $BF$ gauge theory can be
obtained from the action which is the generalized second Chern
class with gauge group $G$. Particularly when $G$ is taken as
$SL(2,C)$ in four dimensions, general relativity with cosmological
constant can be derived by constraining the topological $BF$
theory.
\end{abstract}



\maketitle

%%%%%%%%%%%%%%%%%%%%%%%%%%%%%
% Generalized p-forms
\newcommand{\GForm}[2]{\buildrel\scriptstyle {#1} \over {\bf #2}}
% Example: $\GForm{p}{a}$
%%%%%%%%%%%%%%%%%%%%%%%%%%%%%
% Regular p-forms
\newcommand{\Form}[2]{\buildrel\scriptstyle {#1} \over #2}
% Example: $\Form{p}{\alpha}$
%%%%%%%%%%%%%%%%%%%%%%%%%%%%
% GDC-d
\newcommand{\Gd}{{\bf d}}
% Example: \d
%%%%%%%%%%%%%%%%%%%%%%%%%%%%

%%%%%%%%%%%%%%%%%%%%%%%%%%%%

\section{Introduction}
In the last few years one of the most important progress in
quantum gravity is that deep relations between gravity and the
gauge fields have been further disclosed. In string theory one
seminal work is the proposed $AdS/CFT$
correspondence\cite{Aharony:1999ti}, which implies there might
exist a general duality between gravity theory and Yang-Mills
theory.  On the side of loop quantum gravity, the replacing of
geometrodynamics by connection dynamics has also shed light on the
analogy of gravitational fields and gauge
fields\cite{Ashtekar:hf}. In particular, there has been recently
much interest in the relations between topological field theory
and gravity\cite{CraneYetter,linking}. Besides the well known fact
that three dimensional gravity is simply a topological field
theory without local degrees of freedom\cite{Witten}, recent
progress shows that even in higher dimensions Einstein's general
relativity and supergravity in Ashtekar formalism may also be
written as topological field theories with extra
constraints\cite{CDJM,CDJ,super-ezawa,Smolin,LingSmolin}\footnote{This
idea is similar to that of MacDowell-Mansouri \cite{MM,Mansouri}
in which general relativity is obtained by breaking the $SO(3,2)$
symmetry of a topological field theory down to $SO(3,1)$. The
earlier work to present general relativity as a constrained
topological field theory can also be seen in\cite{Plebanski}}.
Since no pre-exiting metric or other geometric structure of
spacetime is needed in the context of topological field theory,
the advantages of taking this framework as a starting point to
explore the background-independent quantum theory of gravity have
been revealed in the context of loop quantum gravity from many
aspects. Two remarkable programmes are ``state sum model''
advocated by Crane and Yetter\cite{CraneYetter} and ``spin foam
model'' by Rovelli and Reisenberger\cite{foam}.  In both
programmes such connections between gravity and topological field
theories allow one to apply the elegant quantization methods in
topological quantum field theory to quantum gravity. Furthermore,
this formulation provides several features to realize the
holographic principle proposed by 'tHooft and Susskind at quantum
mechanical level\cite{Smolin,LingSmolin}.

Two typical sorts of topological field theories are Chern-Simons
theory and $BF$ theory\cite{Horowitz,BBRT}. Both of them are of
the Schwarz type topological field theories and have important
applications in quantum gravity. In particular one finds once
Einstein-Hilbert action is written as a $BF$ theory with extra
constraints, the behavior of gravitational field on the boundary
of spacetime can be described by Chern-Simons theories after
imposing appropriate boundary conditions. This interesting
intersection raises the question of whether these two kinds of
topological field theories have closer geometric relations.

In this paper, we propose an answer to this question by using a
``{\it Generalized Differential Calculus}'' (see
Appendix)\cite{Sparling,NR2001}. The main assumption is to
generalize the ordinary differential $p-$form to an ordered pair
of $p$ and $(p+1)-$form, and then treat the gauge fields and their
topological properties in the framework of Generalized
Differential Calculus. Using this Generalized Differential
Calculus, we first establish the generalized Chern-Weil
homomormism for generalized curvature invariant polynomials in
general even dimensional manifolds (Section 2). Then we obtain the
generalized Chern-Simons term for BF theory in four dimensions.
This leads to a close relation between these two topological field
theories of the Schwarz type. We also re-derive the geometric
properties for both $P(M^4, G)$ and pseudo-Riemannian spacetime
manifolds from the generalized topological field theory of BF type
(Sections 3 and 4). In other words, we obtain both BF gauge
theories or BF gravity without matter from the action as the
generalized second Chern class with gauge group $G$ or $SL(2,C)$
respectively. As a consequence, we find that GR with cosmological
constant in the absence of matter can be derived by a constrained
topological field theory (Section 5) related to holographic
formulation \cite{Smolin}.

\section{Chern-Weil Homomormism in Generalized Differential Calculus}

Let us consider a principle bundle $P(M, G)$ and introduce the
generalized gauge fields in Generalized Differential Calculus on
it. For the semisimple gauge group $G$ with Lie algebra ${ g}$, a
generalized $g$-valued connection 1-form field ${\cal A}$  is
defined as a $g$-valued pairing of a 1-form and a 2-form
\begin{equation}
{\cal A}=( A^{p}, B^{p} ) T_{p} =(A, B),\quad T_p
\in g, \label{1}
\end{equation}
where ${A} $ is the ordinary $g$-valued
connection 1-form and $B$  the $g$-valued 2-form
which is assumed as gauge covariant under the
gauge transformations in order to introduce a
generalized gauge covariant curvature ${\cal F}$
\begin{eqnarray}
 {\cal F} &=&\Gd{\cal A} + {\cal A} \wedge {\cal A}\nonumber\\
 &=& \left( dA+A\wedge
 A+k B, \quad  d B+
 A\wedge B -B \wedge A \right) \nonumber\\
 &=&(F+ k B,~ D B).
\end{eqnarray}
It satisfies the generalized Bianchi identity:
\begin{eqnarray}\label{GBI}
 {\cal D} {\cal F} &=& \Gd {\cal F}
 + {\cal A}\wedge
 {\cal F} - {\cal F} \wedge {\cal A}
 \nonumber \\
 &=&( DF, ~D^2 B) \equiv 0 .
\end{eqnarray}

 In order to consider the topological invariants
in the framework of this Generalized Differential Calculus, let us
first briefly remind the properties of the curvature invariant
symmetric polynomials on $P(M,G)$.

Taking a connection 1-form $A$ on the bundle, the curvature 2-form
is $F= dA + A \wedge A$. The curvature invariant  symmetric
polynomial, say, for simplicity, $P(F^m)$ is a $2m$-form on $M$
\begin{equation}
P(F^m) = P(\underbrace{F,\cdots, F}_m)
\end{equation}
satisfying (a) $P(F)$ is  closed, i.e.,
\begin{eqnarray}
dP(F^m) = 0,
\end{eqnarray}
(b) $P(F^m)$ has topologically invariant integrals.  Namely, it
satisfies the Chern-Weil homomormism formula:
\begin{eqnarray}\label{CW} P(F^m_1) -
P(F^m_0) = dQ(A_0, A_1),
\end{eqnarray} where
\begin{equation}
 Q(A_0, A_1) = \int\limits^1_0
P(A_1- A_0, F^{m-1}_t) dt,
\end{equation}
where $A_0$ and $A_1$ are two connection 1-forms, $F_0$ and $F_1$
the corresponding curvature 2-forms,
\begin{equation}
A_t = A_0 + t\eta,~~\eta = A_1 - A_0,~~~ (0\leq t
\leq 1),
\end{equation}
 the interpolation between $A_0$ and
$A_1$,
\begin{eqnarray}
F_t = dA_t + A_t \wedge A_t .
\end{eqnarray}
Since %
$P(F^m_1)$ and $P(F^m_0)$ differ by  an exact form, their
integrals over manifolds without boundary give the same results
and $ Q(A_0, A_1)$ is called the secondary topological class.

For the generalized connection ${\cal A}$ and curvature ${\cal F}$
in Generalized Differential Calculus, it can be proved that the
generalized curvature invariant symmetric polynomial, say, for
simplicity, ${\cal P}({\cal F}^m)$ also satisfies the similar
closed condition and the generalized Chern-Weil homomormism
formula:
\begin{eqnarray}
(i) \qquad&& \Gd{\cal P}({\cal F}^m)=0 , \\
(ii) \qquad&& {\cal P}({\cal F}^m_1)-{\cal P}({\cal F}^m_0)=
\Gd{\cal Q}({\cal A}_0, {\cal A}_1). \label{cs}
\end{eqnarray}

Let us now sketch the proof. For proving (i), it is a
straightforward consequence by using the generalized Bianchi
identity (\ref{GBI}).

To prove (ii), let us take  two distinct generalized connections
${\cal A}_0$, ${\cal A}_1$ and the corresponding curvatures ${\cal
F}_0$, ${\cal F}_1$ on the bundle. Let
\begin{eqnarray}
{\cal A}_t={\cal A}_0+t {\mbox {\boldmath
$\eta$}}, \qquad {\mbox {\boldmath $\eta$}}={\cal
A}_1-{\cal A}_0, ~~~ (0\leq t \leq 1),
\end{eqnarray}
and the corresponding curvature is
\begin{eqnarray}
{\cal F}_t&=&\Gd {\cal A}_t+{\cal A}_t\wedge
{\cal A}_t.
\end{eqnarray}
It is easy to see that
\begin{eqnarray}
{d \over dt} {\cal F}_t=\Gd {\bm \eta}+{\cal A}_t\wedge {\bm
\eta}+{\bm \eta}\wedge {\cal A}_t\equiv {\cal D}_t{\bm \eta}.
\end{eqnarray}
 Hence
\begin{eqnarray}
{d\over dt} {\cal P}({\cal F}_t^m)&=&m {\cal P} ({\cal D}_t{\mbox
{\boldmath $\eta$}},{\cal F}_t^{m-1}) \nonumber\\
=m {\cal D}_t {\cal P}({\mbox {\boldmath $\eta$}},{\cal
F}_t^{m-1}) &=& m \Gd {\cal P}({\mbox {\boldmath $\eta$}}, {\cal
F}_t^{m-1}),
\end{eqnarray}
Thus
\begin{eqnarray}
{\cal P}({\cal F}^m_1)-{\cal P}({\cal F}^m_0)
 &=&\Gd \int_0^1
{\cal P} ({\cal A}_1-{\cal A}_0 ,{\cal F}^{m-1}_t) dt \nonumber\\
 &=&\Gd
{\cal Q} ({\cal A}_0, {\cal A}_1).
\end{eqnarray}
This shows that the generalized characteristic polynomials with
respect to different connections only differ by an exact form in
Generalized Differential Calculus. Namely, they are also
homomormism. ${\cal Q}({\cal A}_1,{\cal A}_0)$ is called the
generalized  Chern-Simons secondary class.

Thus, we have established the generalized
Chern-Weil homomormism for generalized curvature
invariant polynomials in any even dimensional
manifolds. But, their topological meaning should
be as same as before.

\section{generalized Chern-Simons Term for BF theory}

Consider  an action on the base manifold of
$P(M^4, G)$ of the form :
\begin{eqnarray}\label{a}
{\cal S}_T=\int_{M^4}{\cal L}_T=\int_{M^4} Tr ({\cal F}\wedge
{\cal F}).
\end{eqnarray}
The Lagrangian 4-form ${\cal L}_T$ can be given
by taking ${\cal A}_1={\cal A}$ and ${\cal
A}_0=0$ in (\ref{cs}), then
\begin{eqnarray}\label{GCW}
Tr ({\cal F}\wedge {\cal F})=\Gd{\cal Q}_{CS},
\end{eqnarray}
${\cal Q}_{CS}$ is the generalized local Chern-Simons 3-form,
i.e., the  pairing of a 3-form and a 4-form
\begin{eqnarray}\label{CS}{\cal Q}_{CS}
&=& Tr ({\cal A} \wedge {\cal F}-{\textstyle{1\over3}}
{\cal A} \wedge {\cal A} \wedge {\cal A})\nonumber\\
&=& ( Tr (A \wedge F - {\textstyle{1\over3}} A \wedge A \wedge A
+k A\wedge B),  \nonumber\\
&& \qquad ~ Tr( A \wedge DB+B\wedge F+k B\wedge B) ).
\end{eqnarray}
In the pairing,  the 3-form is the usual Chern-Simons term up to a
$k Tr(A\wedge B)$ term.

On the other hand, the generalized Lagrangian 4-form in (\ref{a})
is a pairing of a 4-form and a 5-form:
\begin{eqnarray}
{\cal L}_{T}%
&=& Tr( (F\wedge F + 2k B\wedge F+k^2 B \wedge B), \nonumber\\
&&\qquad 2\, (F \wedge DB+k B\wedge DB) ).
\end{eqnarray}
Using the Bianchi identity, we can rearrange the
5-form so that
\begin{eqnarray}
{\cal L}_{T}&=& Tr((F \wedge F + 2k B\wedge F+k^2 B\wedge
B), \nonumber\\
&&\qquad ~ d ( B \wedge F+k B\wedge B ) ).
\end{eqnarray}
The first term is just the BF Lagrangian up to an
$F\wedge F$ term, the second term is a total
derivative of the BF Lagrangian.

Thus the pairing of the action (\ref{a}) shows a relation between
two types of topological field theories, the Chern-Simons type and
the BF type in four dimensions :
\begin{eqnarray}
{\cal S}_{T}[{\cal A}]
&=&\int_{M^4} {\cal L}_T=\int_{M^4} \Gd{\cal Q}_{CS} \nonumber\\
 &=& \int_{M^4}  Tr \left( F\wedge F + 2 k B \wedge F + k^2 B\wedge
 B \right).
 \end{eqnarray}

We can obtain the field equations by varying the
Lagrangian with respect to the generalized gauge
potentials, i.e., the $g$-valued generalized
connection
 1-form. With these generalized gauge
potentials fixed at the boundary, the field
equations are
\begin{eqnarray}
&& D(F + 2 k B) =0,\label{11} \\
&& k (F + k B)=0 .\label{12}
\end{eqnarray}
The second equation gives  $F=-k B$ which can
be substituted %
into the first equation and leads to the Bianchi identity. Note
that we start up with any 2-form $B$ in the generalized
connection. Then with the action given by the generalized second
Chern class, it gives $B=-(1/k) F$. Therefore, the action
(\ref{a}) does not give any dynamics.

However, it should be noted that with $k=-1$ all geometrical
properties of the bundle $P(M^4, G)$ are re-derived from a
generalized topological field theory of the BF type with a
generalized Chern-Simons term associated in Generalized
Differential Calculus.

\section{Generalized Connection with $SL(2,C)$ Gauge Group on
Pseudo-Riemannian Manifold $M^4$}

Let us consider the tangent bundle $T(M^4)\simeq $ $ P(M^4,
SL(2,C))$ on the base manifold $(M^4, g)$ as the pseudo-Riemannian
spacetime manifold with signature $sign(g)=-2$. The $sl(2,C)$
algebraic relation reads {\footnote{The upper-case Latin letters
$A,B,...=0,1$ denote two component spinor indices, which are
raised and lowered with the constant symplectic spinors
$\epsilon_{AB}=-\epsilon_{BA}$ together with its inverse and their
conjugates according to the conventions
$\epsilon_{01}=\epsilon^{01}=+1$,
$\lambda^A:=\epsilon^{AB}\lambda_B$, $\mu_B:=\mu^A\epsilon_{AB}$
\cite{PR}.}}:
\begin{eqnarray}
\left[ M_{AB}, M_{CD} \right] &=&
      \epsilon_{C(A} M_{B)D}
     +\epsilon_{D(A} M_{B)C} ,
\end{eqnarray}
where $\epsilon_{C(A} M_{B)D} =\textstyle{1\over2} (\epsilon_{CA}
M_{BD} +\epsilon_{CB} M_{AD})$. The Cartan-Killing metric
$\eta_{pq}={\rm{diag}}(\eta_{(AB)(MN)})$ is given by
\begin{eqnarray}
\eta_{(AB)(MN)}&=& \textstyle{1\over2}
(\epsilon_{AM}\epsilon_{BN}+\epsilon_{AN}\epsilon_{BM}).
\end{eqnarray}

Since $SL(2,C)$, the covering of the Lorentz group $SO(3,1)$, is
the  gauge group of the bundle, we may introduce an
$sl(2,C)$-valued generalized connection 1-form in the framework of
Generalized Differential Calculus
\begin{equation}
{\cal A}=(\omega^{AB}, B^{AB} ) M_{AB} , \label{71}
\end{equation}
where $\omega^{AB}$ is an ordinary $sl(2,C)$-valued connection
1-form on the bundle and $B^{AB}$ is an $SL(2,C)$-gauge covariant
2-form. Given the connection ${\cal A}$, the generalized curvature
(${\cal F}={\cal F}^p T_p={\cal F}^{AB} M_{AB}$) is given by
\begin{equation}
{\cal F}^{AB}=(R^{AB}+ k B^{AB},~ D B^{AB}) .
\end{equation}
where $R^{AB}=d \omega^{AB}+ \omega^A{}_C\wedge \omega^{CB}$ is
the $SL(2,C)$ curvature 2-form. The generalized Bianchi identity
is given by
\begin{equation}
 {\cal D} {\cal F}^{AB} =
( DR^{AB}, D^2 B^{AB})\equiv 0 . \label{Bianchi-SL2C}
\end{equation}
A simple generalized Lagrangian  4-form in (\ref{a}) using this
connection $\cal A$ is
\begin{eqnarray}
&&{\cal S}_{{SL(2,C)}}[{\cal A}]=\int_{M^4} {\cal L}_{{SL(2,C)}}
=\int_{M^4} \, {\cal F}^{AB}\wedge {\cal F}_{AB} \nonumber\\
&=& \int_{M^4} \, R^{AB} \wedge R_{AB} + 2k R^{AB}\wedge
B_{AB}+k^2 B^{AB} \wedge B_{AB} . \nonumber\\
\label{SL2C-action}
\end{eqnarray}
The field equations are obtained by varying the Lagrangian with
respect to the $sl(2,C)$-valued generalized connection 1-form with
fixed value at the boundary. This leads to the field equations
\begin{eqnarray}
&& D(R^{AB} + 2 k B^{AB}) =0,\label{81} \\
&& k (R^{AB}+ k B^{AB})=0.\label{82}
\end{eqnarray}
The second equation gives  $R^{AB}=-k B^{AB}$, which can be
substituted into the first and leads to the Bianchi identity.
Therefore, as expected, the action ${\cal S}_{{SL(2,C)}}$ does not
give any dynamics. Moreover, when $k=-1$, all properties in the
pseudo-Riemannian geometry on $(M^4, g)$ are recovered by the
generalized topological field theories of BF type in four
dimensions.

\section{Gravity as a generalized Topological Field Theory}

Let us now consider pure gravity in four
dimensions. As in the previous section, the
spacetime manifold $(M^4, g)$ is
 pseudo-Riemannian with signature $sign(g)=-2$
and the gauge group now is $SL(2, C)$.

 Note that  the
4-form Bianchi identity (\ref{Bianchi-SL2C}), $D^2 B^{AB}=0$,
looks similar to the identity $D^2(e^{AA'}\wedge e^{B}{}_{A'})=0$,
where $e^{AA'}$ is the frame 1-form. Thus we may introduce an
ansatz $B^{AB}= l^{-2}~ e^{AA'}\wedge e^B{}_{A'}$, where $l$ is a
dimensional constant. The $sl(2,C)$-valued generalized connection
1-form (\ref{71}) now becomes{\footnote{The formulation can be
written with purely unprimed spinors by defining spinor 1-forms
$\varphi^A=e^{A0'}$ and $\chi^A=e^{A1'}$ \cite{TJ}. In terms of
these spinor 1-forms, the purely unprimed $sl(2,C)$-valued
generalized connection 1-form
 is ${\cal A}^+=\left(\omega^{AB}, ~~(2 / l^2)~
\chi^{(A} \wedge \varphi^{B)} \right) M_{AB}$ .}}
\begin{equation}
{\cal A}=(\omega^{AB}, ~~{1\over l^2}
e^{AA'}\wedge e^B{}_{A'} ) M_{AB} +c.c.
\label{61}
\end{equation}
We shall show that given the above generalized connection 1-form
(\ref{61}), the action of generalized topological field theory
type (\ref{SL2C-action}) becomes the Hilbert action with
cosmological constant plus an ordinary topological term. If we
vary this generalized topological field theory action with respect
to the introduced new variable $e^{AA'}$, it yields the Einstein
equation with a cosmological constant term.

Using the generalized connection 1-form (\ref{61}), the
generalized curvature is given by
\begin{equation}
{\cal F} =(R^{AB} + {k\over l^2} e^{AA'}\wedge
 e^B{}_{A'},~~
 {1\over l^2}D (e^{AA'}\wedge e^B{}_{A'}) ) M_{AB} +c.c.
\end{equation}
In terms of this generalized curvature, the action
(\ref{SL2C-action}), with a change of variational variables from
$B^{AB}$ to $e^{AA'}$,
\begin{eqnarray}
&& {\cal
S}[\omega^{AB},\omega^{A'B'},e^{AA'}]=\int_{M^4}
Tr({\cal F}
\wedge {\cal F})+c.c. \nonumber\\
&=& \int_{M^4}\,( R^{AB} \wedge R_{AB} + {2k\over l^2}
R^{AB}\wedge e^{AA'}\wedge e^B{}_{A'} \nonumber\\
 &&+ {k^2\over l^4}
e^{AA'}\wedge e^B{}_{A'} \wedge e_{A}{}^{C'}\wedge e_{BC'} )
+c.c.,
\end{eqnarray}
 gives the Einstein-Hilbert action with the cosmological
constant plus a topological term. Namely, General Relativity in
the absence of matter is formulated as a generalized topological
field theory.

By varying with respect to $\omega^{AB}$, we
obtain
\begin{equation}
D(e^{AA'}\wedge e^{B}{}_{A'})=0 ,\label{FE1}
\end{equation}
which gives the equation for  torsion-free. While
by varying with respect to $e^{AA'}$, we obtain
\begin{equation}
R^{AB}\wedge e_B{}^{A'}+{k\over l^2} e^{AB'}\wedge e^{B}{}_{B'}
\wedge e_B{}^{A'}+c.c.=0 ,\label{FE2}
\end{equation}
which is the Einstein equation with a cosmological constant
$\Lambda $ if we set $\Lambda$ by $\Lambda ={k\over l^2}$.

Alternatively,  we can consider adding a constraint on $B^{AB}$ as
in \cite{Smolin}
\begin{eqnarray}
&&{\cal S}[{\cal A}^p,\lambda_{AB},e^{AA'}] =\int_{M^4}\,
{\cal F}^{AB}\wedge {\cal F}_{AB} \nonumber\\
&&+\lambda_{AB}
\wedge ({1\over l^2} e^{AA'}
\wedge e^{B}{}_{A'}-B^{AB}) + c.c.\nonumber\\
&=& \int_{M^4}\, R^{AB} \wedge R_{AB} + 2k
R^{AB}\wedge B_{AB}+k^2
B^{AB}\wedge B_{AB}  \nonumber\\
 &&+ \lambda_{AB} \wedge
( {1\over l^2} e^{AA'}\wedge e^{B}{}_{A'}-B^{AB}) + c.c.,
\end{eqnarray}
where $\lambda_{AB}$ is the Lagrangian multiplier. The variational
principle leads to the same equation as (\ref{FE2}).

In the last part of this section we note that as the cosmological
constant, namely the constant $k$ goes to zero, the equation
(\ref{FE2}) does not necessarily hold any more. It implies our
generalization perhaps is only valid for the cases with non-zero
cosmological constant.

\section{Discussion}

In this paper, we have generalized the Chern-Weil  homomormism in
Generalized Differential Calculus, associated the generalized
Chern-Simons Lagrangian to BF theories and re-derived geometrical
properties of $P(M^4, G)$ from a generalized topological field
theory of the BF type on four dimensions. We have also recovered
the properties of the pseudo-Riemannian manifold $M^4$ from a
generalized topological field theory of the BF type and
reformulated GR in the absence of matter as either a generalized
topological field theory or a constrained one.

For General Relativity, our approach starts with an $sl(2,C)$-
valued generalized connection which includes the 2-form $B$
fields. In a sense, this approach is similar to that of
MacDowell-Mansouri \cite{MM}, in which General Relativity is found
as a consequence of breaking the $Sp(4)$ symmetry of a topological
field theory down to Lorentz group's covering group $SL(2,C)$ and
introduce the tetrad 1-form fields $e^{AA'}$ to parameterize the
coset $Sp(4)/SL(2,C)$. However, our approach differs from theirs.
Instead of breaking the symmetry of a topological field theory, we
start with the Lorentz group and redefined a new generalized
$sl(2,C)$-valued connection in Generalized Differential Calculus.
This directly leads to a BF theory.

In order to include the matter, there might be at least two
possibilities. The first may link with the Kaluza-Klein formalism,
since the gauge theories can be formulated as Kaluza-Klein
theories on Minkowski spacetime. Therefore,  it might be possible
to deal with gauge theories as generalized topological field
theories in our approach. Of course, there should be certain
restriction to the dimensions of the gauge groups. Furthermore,
this approach might be generalized to fermions and Higgs
\cite{liu}. On the other hand, to generalize it to supergravity
might be another possibility. For instance, the Generalized
Differential Calculus may be generalized to supersymmetric cases.
Then, supergravity can be obtained by gauging the $OSp(1,4)$ group
with the generalized connection.

It is interesting to see that the present formulation only works
for 4-dimensional BF theories with $B$ a 2-form field. On one
hand, it is reasonable to establish the relation between the
Donaldson-Witten invariants in four dimensions and the topological
field theory such as the BF type. On the other hand, however,
especially for the GR with cosmological constant, it seems amazing
that the dimensions of our nature is also four. If this
formulation could not be generalized to arbitrary higher
dimensions, whether  this dimension four has more profound meaning
rather than just a coincidence. This question has to be left for
further study and inspiration.

\section*{Appendix: Generalized Differential Calculus}

A generalized $p$-form \cite{Sparling}\cite{NR2001},
$\GForm{p}{a}$, is defined to be an ordered pair of an ordinary
$p$-form $\Form{p}{\alpha}$ and an ordinary $(p+1)$-form
$\Form{p+1}{\alpha}$ on an n-dimensional manifold $M$, that is
\begin{equation}
\GForm{p}{a} \equiv (\Form{p}{\alpha},\Form{p+1}{\alpha}) \in
\Lambda^p \times \Lambda^{p+1} ,
\end{equation}
where $-1 \leq p \leq n$. The minus one-form is defined to be an
ordered pair
\begin{equation}
\GForm{-1}{a}\equiv (0, \Form{0}{\alpha}) ,
\end{equation}
where $\Form{0}{\alpha}$ is a function on $M$. The product and
derivatives are defined by
\begin{equation}
\GForm{p}{a} \wedge \GForm{q}{b}\equiv ( \Form{p}{\alpha}\wedge
\Form{q}{\beta}, \Form{p}{\alpha} \wedge \Form{q+1}{\beta}+(-1)^q
\Form{p+1}{\alpha}\wedge \Form{q}{\beta}) ,
\end{equation}
\begin{equation}
\Gd \GForm{p}{a} \equiv ( d \Form{p}{\alpha}+(-1)^{p+1} k
\Form{p+1}{\alpha}, d \Form{p+1}{\alpha}) ,
\end{equation}
where $k$ is a constant. These exterior products
and derivatives of generalized forms satisfy the
standard rules of exterior algebra
\begin{equation}
\GForm{p}{a} \wedge \GForm{q}{b} = (-1)^{pq} \GForm{q}{b} \wedge
\GForm{p}{a} ,
\end{equation}
\begin{equation}
\Gd (\GForm{p}{a} \wedge \GForm{q}{b}) = \Gd \GForm{p}{a} \wedge
\GForm{q}{b} + (-1)^{p} \GForm{p}{a} \wedge \Gd \GForm{q}{b} ,
\end{equation}
and ${\bf d}^2=0$.

For a generalized p-form $\GForm{p}{a}=
(\Form{p}{\alpha},\Form{p+1}{\alpha})$, the
integration on $M^p$ can be defined as usual by
\begin{equation}
\int_{M^p} \GForm{p}{a}=\int_{M^p}
(\Form{p}{\alpha},\Form{p+1}{\alpha}) =
\int_{M^p} \Form{p}{\alpha} .
\end{equation}

\begin{acknowledgments}
We would like to thank Professors Y.F. Liu, Q.K. Lu, K. Wu, Z. Xu
and M. Yu for valuable discussions. RST thanks Prof S. Carlip for
helpful discussions. This project is in part supported by NNSFC
under Grants Nos. 90103004, 10175070, 10047004, 19835040 and also
by NKBRSF G19990754.
\end{acknowledgments}

\end{document}